# Electric Field-Induced Phase Transitions and Hysteresis in Ferroelectric $HfO_2$ Captured with Machine Learning Potential

*Po-Yen Chen[a*], Teruyasu Mizoguchi[a,b*].*

AFFILIATIONS

[a]Department of Materials Engineering, the University of Tokyo, Tokyo, Japan

[b]Institute of Industrial Science, the University of Tokyo, Tokyo, Japan.

AUTHOR INFORMATION

**Corresponding Author**

poyen@iis.u-tokyo.ac.jp, teru@iis.u-tokyo.ac.jp

ABSTRACT

Electric field-induced studies, including phase transition and polarization hysteresis, for ferroelectric $HfO_2$ at the atomic scale are critical since they can largely affect its application in ferroelectric and dielectric devices. However, conventional first-principles approaches are computationally limited in capturing large-scale atomic dynamics under realistic field conditions. Here, to enable electric-field-driven molecular dynamics simulations, we develop a machine learning potential (MLP) tailored for $HfO_2$, coupled with an in-situ Born effective charge (BEC) model. This framework enables us to capture key phenomena, including field-induced phase transitions, polarization switching, and strain-dependent dielectric responses, with high fidelity and computational efficiency. Notably, we reproduce hysteresis loops and phase transition barriers consistent with AIMD results and reveal possible electric-field-induced polarization activation in the monoclinic phase. Our approach offers a scalable and transferable tool for atomistic exploration of functional oxides and paves the way for data-driven design of ferroelectric devices.

1. Introduction

HfO$_2$ has become a cornerstone material in modern semiconductor technology, widely adopted as a dielectric in ferroelectric capacitors (FeCaps), metal-oxide-semiconductor field-effect transistors (MOSFETs), and complementary metal-oxide-semiconductor (CMOS) devices[1-5]. More recently, the discovery of ferroelectricity in doped or structurally engineered HfO$_2$ has led to the development of ferroelectric field-effect transistors (FeFETs)[6], positioning HfO$_2$ as a key candidate for next-generation non-volatile memory and energy-efficient switching devices. To fully exploit these properties, a deep understanding of the material's electric field response at the atomic scale, particularly under device-relevant conditions, is essential. While ab initio molecular dynamics (AIMD) based on density functional theory (DFT) has been widely used to study electric-field-induced structural changes[7, 8], its high computational cost restricts simulations to small spatial and temporal domains. As a result, critical processes such as large-scale polarization switching, domain wall motion, and phase transitions in realistic environments remain out of reach for AIMD alone. This limitation underscores the urgent need for scalable and accurate modeling frameworks that can capture both atomic-level dynamics and electronic responses under external electric fields.

To overcome these limitations, machine learning potentials (MLPs) have recently emerged as promising alternatives[9-11]. MLPs are data-driven interatomic models trained on high-fidelity first-principles datasets, enabling the prediction of atomic forces and energies with near-DFT accuracy. They allow for efficient and scalable MD simulations that go beyond the capabilities of conventional empirical force fields[12], and, consequently, enable the reproduction and exploration of properties with complicated material structures[13-15]. Despite these advantages, most existing MLPs are trained primarily on structural and energetic data, without explicitly incorporating electronic information. This restricts their applicability to systems where electric field effects play a central role.

To address this challenge, the incorporation of Born effective charges (BECs) plays a pivotal role. BECs quantify the change in polarization resulting from atomic displacements and enable the evaluation of electric-field-induced atomic forces, thus serving as a critical link between atomic-scale dynamics and macroscopic dielectric responses[16]. Unlike static charges such as Bader charges, BECs are directly related to variations in polarization or dipole moment, which is typically inferred indirectly from experimental data such as phonon mode strengths[17]. This hybrid approach captures electric-field-driven phenomena such as phase transitions, hysteresis loops, and dielectric responses—bridging the gap between first-principles accuracy and realistic simulation scales.

Several studies have incorporated BECs into external electric force calculations, demonstrating their feasibility in MD simulations[18-21]. Especially, Falletta *et al.*[19] developed a machine learning (ML) model that predicts the electric enthalpy, from which BECs, atomic forces, and polarizations are obtained via automatic differentiation, enabling highly efficient MLMD simulations under external electric fields. While this framework offers significant advantages in scalability and response-property consistency, it requires training data that includes DFT calculations both in the absence and in the presence of a small electric field applied along each Cartesian direction, which can be relatively time-consuming and computationally demanding during data preparation.

In contrast, we considered that the approach directly incorporates precomputed BECs into the force evaluation, implementing electric-field-coupled MD simulations more straightforwardly and practically effectively. To enable fast and instantaneous BEC evaluation during MD simulations, we employ the Equivar model, an equivariant graph convolutional neural network (GCNN) framework proposed by Kutana *et al.*[22] Its pretrained version of this model, BM1, has demonstrated reliable performance across a range of materials, including perovskite oxides, $Li_3PO_4$, and $ZrO_2$. This advancement provides a new pathway for incorporating electronic response information into machine-learning-based simulations without the need for repeated first-principles calculations. In this study, we aim to combine an MLP with BEC predictions to efficiently simulate the electric field response of $HfO_2$. Specifically, we investigate key properties such as polarization hysteresis, electric-field-induced phase stability, and the strain-dependent dielectric constant. This approach not only enables the accurate and efficient modeling of field-driven phenomena in $HfO_2$ but also provides valuable insights into its functional behavior under realistic device operating conditions. We believe that our findings will contribute to the fundamental understanding of ferroelectric responses in $HfO_2$ and support the development of advanced $HfO_2$-based components in next-generation semiconductor devices.

2. Results and discussion

To ensure high accuracy in modeling the $HfO_2$ system, we fine-tuned the universal machine learning potential MACE-MP-0[23, 24] using a comprehensive dataset generated from first-principles MD simulations with 3,478 data. This dataset encompasses six distinct crystallographic phases of $HfO_2$, including $Fm\bar{3}m$, $P2_1/c$, $P4_2nmc$, $Pbca$, $Pca2_1$, and $Pnma$, sampled over a wide temperature range from 1 K to 4000 K. The detailed procedures for dataset generation and model training are described in the Methods section.

The predictive accuracy of the resulting HfO$_2$-specific MLP is illustrated in Figure 1a and 1b, which presents the mean absolute errors (MAEs) of 2.14 meV/atom for total energies and 30.2 meV/Å for atomic forces, evaluated 869 test data across six distinct crystalline phases of HfO$_2$ under a range of temperatures. These results indicate our finetuned model has excellent agreement with reference DFT values. For further comparison, we evaluated the performance of several other published universal MLPs, including MACE-MP-0[23], SevernNet-0[25], MatterSim[26], and CHGNET[27], on the same HfO$_2$ test dataset, as shown in Table 1. Our fine-tuned model outperforms these existing universal models in both energy and force predictions, validating the effectiveness of system-specific refinement and demonstrating the high accuracy and transferability of our approach for HfO$_2$ simulations.

Since the ferroelectric orthorhombic Pca2$_1$ phase is the most responsive to external electric fields, our analysis in this study primarily focuses on this phase. To validate the structural reliability of our model, we compared the lattice parameters of the optimized Pca2$_1$ structure obtained from fine-tuned MACE with both DFT calculations and experimental data. The MACE-optimized structure yields lattice constants of a = 5.01 Å, b = 5.03 Å, and c = 5.21 Å, which are in excellent agreement with the DFT values (a = 5.00 Å, b = 5.03 Å, c = 5.21 Å), demonstrating the model's ability to accurately reproduce the ferroelectric crystal structure. As pure ferroelectric Pca2$_1$ HfO$_2$ does not exist under ambient conditions, experimental lattice parameters are often influenced by substrate constraints and dopant effects. In the work by Yun et al.[28], for example, HfO$_2$ doped with 5% Y and grown on an LSMO substrate exhibited lattice constants of a = b = 5.07 Å and c = 5.20 Å, which are comparable to our simulation results. This agreement between our optimized structure and both theoretical and experimental lattice constants supports the validity of our model and its suitability for further investigation of field-induced ferroelectric behavior.

As a preliminary step, we investigated the size dependence of the phase transition from the $Pca2_1$ phase to the high-temperature tetragonal $P4_2nmc$ phase. This transition was characterized by tracking the changes in lattice parameters and the absolute displacement of oxygen atoms along the [010] direction, as illustrated in Figure 2a. The results for the 8×8×8 supercell (corresponding to approximately 4.1×4.1×4.2 nm and 6144 atoms) reveal a clear phase transition behavior. Upon heating, a sharp structural change occurs at approximately 1530 K, evidenced by a sudden contraction along the c-axis and a simultaneous expansion along the a-axis. Notably, the lattice constants of the a- and b-axes become equal beyond this point, consistent with the symmetry of the tetragonal phase. Furthermore, the oxygen displacement along the polar direction exhibits a significant reduction, indicating the loss of spontaneous polarization and confirming the transition to the non-polar $P4_2nmc$ phase. The dependence of the phase transition temperature on supercell size is summarized in Figure 2b. As the supercell size increases, the transition temperature systematically rises and eventually converges when the supercell reaches 7×7×7 (corresponding to 4116 atoms). In the previous $HfO_2$ MLP model developed by Wu *et al.*[9], the transition temperature was reported to be approximately 1200 K for a 96-atom supercell and 2000 K for a larger supercell, which is slightly different from our results. Compared to phase transition studies conducted via AIMD using a 2×2×2 supercell (96 atoms), our model yields a similar transition temperature of approximately 1100 K, indicating good agreement and reproducibility with AIMD results[7]. Although AIMD or experimental data for the converged transition temperature of the $Pca2_1$ → $P4_2nmc$ phase transition are limited, Schroeder *et al.* employed DFT-based thermodynamic calculations to estimate the transition temperature of pure $HfO_2$ at around 1400 K by comparing the free energies of the phases[29], further supporting the reliability of our model. Although it is difficult to directly compare our predicted transition temperature with experimental data due to the absence of the pure orthorhombic phase of $HfO_2$ under ambient conditions, a study by Tashiro et al. reports that the transition temperature increases nearly linearly with Y concentration in Y-doped $HfO_2$ thin films (12–18 nm thick) within the 5–7% doping range[30]. By linearly extrapolating this trend toward zero Y concentration, the transition temperature for pure $HfO_2$ can be estimated to be approximately 1573 K, which is close to our predicted value. While such extrapolation should be interpreted with caution, the agreement may lend partial support to the validity of our model.

After validating the predictive performance of our MLP for $HfO_2$, we proceeded to incorporate BECs into the MD simulations to enable the modeling of electric field-induced responses. As a first step, we evaluated the prediction accuracy of the BM1 model for $HfO_2$ systems. Specifically, we compared the BECs predicted by BM1 with those obtained from DFT calculations across six optimized $HfO_2$ phases. The comparison results, presented in Figure 3, demonstrate that the BM1 model exhibits excellent agreement with DFT-derived values, with an average MAE of approximately 0.065 e. Considering that the magnitude of external electric fields commonly employed in simulations is below 20 MV/cm, this level of accuracy corresponds to a force error induced by BECs of less than 13 meV/Å, which is generally acceptable for MD simulations. Since the BM1 training dataset does not include structures under high-pressure conditions, the prediction accuracy of the BECs for high-pressure phases such as Pnma is slightly lower than for other phases. However, the focus of this study is on the ferroelectric $Pca2_1$ phase and its transitions to Pbca and $P2_1/c$. Therefore, the minor deviations in the BEC predictions for the Pnma phase do not significantly affect our overall conclusions. These results confirm the suitability of the BM1 model for accurately describing BECs in HfO$_2$ systems and support its integration into electric-field-enabled MD simulations.

Having confirmed the accuracy of the BEC calculations using the BM1 model, we further employ this model to investigate the phase stability and energy barriers associated with phase transitions in HfO$_2$ under applied electric fields. To this end, we perform nudged elastic band (NEB) calculations[31] for two representative polar–nonpolar phase transitions, including the Pca2$_1$ → Pbca transition and the P2$_1$/c → Pca2$_1$ transition. The Pca2$_1$ → Pbca phase transition involves the migration of oxygen atoms. In particular, for domain wall formation and motion, it has been reported that oxygen migration paths crossing the HfO$_2$ lattice exhibit lower energy barriers compared to non-crossing paths[20, 32-34]. Therefore, investigating this phase transition provides valuable insight into the mechanisms underlying domain wall generation and migration under an applied electric field. In contrast, the P2$_1$/c → Pca2$_1$ transition provides insights into the activation of polarization in HfO$_2$. The detailed method for calculating NEB and its response to electric fields is discussed in the Method section.

The scheme and NEB results are presented in Figures 4a and 4b, respectively. For the Pca2$_1$-Pbca phase transition, the non-polar Pbca phase was used as the reference structure, with its total energy set to zero and atomic displacements defined relative to this configuration. Compared to the Pbca phase, the Pca2$_1$ phase exhibits polarization along the positive b-axis. As a result, under an electric field applied in the positive b direction, the Pca2$_1$ phase becomes more energetically favorable. Besides, the energy barrier between phases is influenced by the direction of the electric field. When a negative electric field is applied, the barrier for the Pca2$_1$ → Pbca transition is reduced, suggesting that the migration of oxygen atoms becomes more favorable. In the study by

Choe et al.[33], it was also demonstrated that the energy barrier for the Pbca → Pca2₁ phase transition decreases with increasing electric field strength, which is consistent with our findings. Besides, the Ma's study also demonstrated this phase transition by utilizing MD simulation[20]. Given that oxygen atom migration plays a key role in domain wall formation and motion in $HfO_2$, this result offers further insight into the field-driven dynamics of domain wall behavior.

For the $P2_1/c$-$Pca2_1$ phase transition, the non-polar $P2_1/c$ phase was selected as the reference structure, defining both the zero-level energy and the origin of ionic displacements. In this transition pathway, the polarization vector of the $Pca2_1$ phase aligns along the negative b-axis. The scheme and NEB results of the phase transition are shown in Figures 5a and 5b, respectively. The $Pca2_1$ phase becomes energetically less favorable under a positive electric field, whereas it is stabilized under a negative electric field. This trend is consistent with previous studies on electric-field-induced phase stability, as demonstrated by Kingsland et al. and Batra et al.[35, 36] Moreover, the energy barrier for the $P2_1/c$ → $Pca2_1$ transition is significantly reduced under a negative electric field, suggesting enhanced polarizability of the $P2_1/c$ phase in strong electric fields and highlighting the possibility of electric-field-induced polarization activation in monoclinic $HfO_2$. While previous studies have rarely observed polarization activation in the $P2_1/c$ phase of $HfO_2$, typically comparing only the total energies of the $P2_1/c$ and $Pca2_1$ phases under an applied field without considering energy barriers[35, 36], our results explicitly demonstrate that the energy barrier is substantially lowered under a strong electric field. This finding suggests the potential for polarization activation in $P2_1/c$-phase $HfO_2$-based materials, particularly those capable of withstanding high breakdown fields.

In the next step, we conducted MD simulations incorporating external electric fields. To enable this, it was necessary to integrate electric forces, which derived from the BECs and the applied electric field, into MLP-based MD simulations. As illustrated in Figure 6, we newly developed the MD framework to explicitly include these additional forces during the force evaluation stage, which have been discussed in various studies[19, 21, 37]. At each MD step, atomic BECs are first predicted using the pretrained BM1 model. These BECs are then used to compute the external electric force on each atom via the relation $F_{ext} = |e|\mathcal{E}_\beta Z^*_{\kappa,\beta\alpha}$. The total force acting on each atom is obtained by summing the MLP-predicted force and the electric field-induced force. These updated forces are subsequently used to propagate atomic positions during the MD simulation. This approach enables efficient and realistic modeling of electric-field-driven phenomena in complex systems while retaining the high computational efficiency of MLPs.

This MD simulation incorporating electric fields can be employed to investigate ferroelectric properties, such as hysteresis behavior and dielectric response. Here, we calculated the polarization–electric field (P–E) hysteresis loop of ferroelectric $Pca2_1$ $HfO_2$ under a single cycle of electric loading. A 2×2×2 supercell—consistent with the cell size used in Fan's AIMD study[7]—was subjected to a varying electric field applied along the polarization direction. The field was ramped sequentially from 0 → -12 → +12 → 0 MV/cm in steps of 3 MV/cm. At each field step, the system was equilibrated for 20 ps, and atomic configurations were recorded every 1 ps. Polarization values were computed using $P_\alpha = (e/\Omega) \sum_{\kappa,\alpha\beta} Z^*_{\kappa,\beta\alpha} u_{\kappa,\alpha}$, where $\Omega$ is the volume of the cell, and the final polarization at each field step was obtained by averaging the data from the last 10 ps (11–20 ps). Figure 7a and 7b show the simulated hysteresis loop and the scheme for polarization switching, respectively. It can be noticed that the simulated hysteresis loop exhibits a remnant polarization of approximately 48 μC/cm² and a coercive field around 10 MV/cm, which is consistent with previous AIMD results. This consistency not only confirms the reproducibility of the ferroelectric hysteresis behavior in HfO₂ but also validates the applicability of the BM1 model for capturing the essential physics of electric-field-driven phenomena in HfO₂ systems. Furthermore, we calculated the dielectric constant by the formula $\varepsilon = \frac{1}{\varepsilon_0}\frac{P}{E} + 1$, and the computed dielectric constant along the polarization direction is 17, while slightly lower than Fan's AIMD results with the value of 27 under tensile strain[7].

In addition to comparing our results with the dielectric constant calculated by Fan's AIMD study[7], we recognized that strain effects may also significantly influence the dielectric properties of HfO₂. Furthermore, it is important to emphasize that strain engineering plays a critical role in practical semiconductor processes, especially in the control and enhancement of ferroelectric polarization[38, 39]. While previous discussions have often focused on how stress affects the phase stability—modulating the ratio between ferroelectric and non-ferroelectric phases—we propose that changes in the intrinsic polarization due to strain may also contribute meaningfully to the dielectric response. Based on this motivation, we incorporated strain into our supercell model to systematically investigate its impact on polarization and dielectric behavior by linear response with small electric fields. We defined the out-of-plane strain direction along with the polarization direction (b-axis), and the a-axis and c-axis are considered as the in-plane strain direction, as shown in Figure 8a. Specifically, we varied the strain from −2% to 2% in both the out-of-plane and equally biaxial in-plane directions under electric fields ranging from −3 MV/cm to 3 MV/cm to evaluate the dielectric constant. Figures 8b and 8c present polarization variation as a function of the applied electric field under out-of-plane and in-plane strain, respectively. For out-of-plane strain shown in Figure 8b, the spontaneous polarization systematically decreases with increasing tensile strain, while it increases when compressive strain is applied. This trend is consistent with the experimental observations by Cheng *et al.*[40] on (111)-oriented $HfO_2$-based thin films, where tensile out-of-plane strain was found to enhance polarization, while compressive strain reduced it under zero epitaxial strain (corresponding to the in-plane strain in our work). Such an agreement reinforces the validity of our simulation in capturing the intrinsic strain–polarization coupling behavior. Additionally, linear regression was performed for each strain condition, and the resulting regression lines are shown as dotted lines. These lines are nearly parallel but exhibit different interceptions corresponding to variations in spontaneous polarization, indicating that the dielectric constant remains relatively insensitive to out-of-plane strain.

In contrast, under equally biaxial in-plane strain shown in Figure 8c, the application of tensile strain leads to a noticeable reduction in spontaneous polarization and an increase in the slope of the regression curves. On the other hand, compressive strain results in a slight increase in polarization, which quickly converges; the corresponding regression curves remain closely aligned, indicating a rapid convergence of dielectric behavior in this regime. In Cheng et al.'s study[40], under tensile epitaxial strain, the polarization (represented by $\Delta d_{111}$ in their research) decreases sharply, while only modest increases are observed under compressive strain, which agrees well with our MD simulation results. These consistent behaviors between our simulations and prior experimental findings underscore the reproducibility of the observed strain-polarization coupling and suggest that it is an intrinsic feature of $HfO_2$-based ferroelectrics.

Although both out-of-plane and in-plane strain simulations demonstrated an increase in spontaneous polarization under compressive strain, experimental observations have reported contrasting trends. Specifically, ferroelectric polarization tends to increase with compressive out-of-plane strain, but decreases under compressive in-plane strain, primarily due to variations in the stability of ferroelectric phases under different strain states[35, 38, 41-44]. This discrepancy indicates that macroscopic ferroelectric behavior is predominantly governed by the strain-induced modulation of the fraction and stability of polar phases, rather than by changes in intrinsic polarization. This interpretation is consistent with the findings of Estandía et al.[45] for $Hf_{0.5}Zr_{0.5}O_2$ (HZO), where it was observed that, despite some HZO thin films exhibiting large intrinsic polarization (represented by $\Delta d_{111}$), their remnant polarization was relatively low. This result suggests that, in practical conditions, the stability of polar phases has a more significant impact on the observed remnant polarization than the intrinsic lattice distortion alone.

Figure 9 summarizes the calculated dielectric constants under both out-of-plane and in-plane equally biaxial strain. Although the regression lines in Figure 8b show only modest changes in slope, the dielectric constant nonetheless increases steadily with increasing out-of-plane strain, which is consistent with previous experimental and computational findings. In the in-plane case, compressive and tensile strains lead to notably different behaviors. As seen in Figure 8c, compressive strain has little effect on the dielectric response, while tensile strain results in a pronounced increase in the dielectric constant. This result also highlights the pronounced influence of tensile in-plane strain on the dielectric constant, offering valuable insight for the design and development of high-permittivity materials.

To enable a better comparison with AIMD simulations, we applied a periodic triangular-wave electric field to $HfO_2$ under 3% equally biaxial tensile in-plane strain. In AIMD studies[7], likely due to computational cost, the periodic electric field was applied only for 200 fs (200 MD steps). However, some experimental studies have reported that the frequency of the applied periodic electric field can influence the coercive field and the shape of the hysteresis loops[46, 47]. To examine this effect, we applied different electric field sweep rates of 200 (0.6k MD steps in total), 20 (2.4k MD steps), 2 (20k MD steps), 0.2 (200k MD steps), and 0.1 kV·cm⁻¹·fs⁻¹ (400k MD steps) within the NPT ensemble to the strained HfO₂. As shown in Figure 10, for sweep rates between 0.1 and 20 kV·cm⁻¹·fs⁻¹, the resulting P-E curves exhibit approximately parallelogram-shaped loops with comparable remnant polarization and linear-regime slopes of polarization change. At the same time, the coercive field systematically decreases as the applied electric field rate decreases and eventually converges to approximately 5 MV/cm. A similar trend has been observed in experimental data of HZO thin films reported by Li *et al.*[47] This suggests that the coercive field of approximately 10 MV/cm reported in AIMD studies[7] may have been overestimated due to the extremely high field frequency employed. By contrast, the P–E curve obtained from only 600 MD steps exhibits a nearly elliptical shape, which deviates significantly from the AIMD results. This discrepancy may arise from the choice of ensemble, as AIMD employed the NVT ensemble[7], which may suppress time-delay effects associated with lattice volume and shape fluctuations. Considering real experimental conditions, Gao *et al.*[46] also observed time-delay effects under high-frequency fields, suggesting that our model is capable of reproducing the essential features observed experimentally.

3. Conclusion

In this study, we developed a highly accurate MLP for HfO₂ and integrated it with a BEC-based Equivariant model to investigate the electric field responses of HfO₂ systems. Our MLP successfully reproduces the field-induced phase transition from the polar Pca2₁ phase to the non-polar P4₂nmc phase, and also captures the size-dependent transition temperature in ferroelectric HfO₂ systems. Furthermore, by leveraging the Equivariant model, we constructed energy diagrams for Pca2₁-to-Pbca and P2₁/c-to-Pca2₁ phase transitions under various electric fields. These results provide insights into the field-dependent stability and energy barrier evolution. Our electric field-driven MD simulations support this hypothesis and additionally reveal that oxygen atom migration may accompany the phase transition process. Moreover, this study successfully reproduces the key electrical properties of $HfO_2$. The hysteresis loop obtained from our simulations closely matches the results from AIMD calculations. The dielectric response under strain also shows good agreement with existing experimental and simulation data, further validating the reliability and predictive capability of our approach. Overall, our work demonstrates that MLP-based MD simulations, combined with effective charge models, offer a powerful and efficient framework for exploring electric field-induced phenomena in ferroelectric materials. This approach not only deepens the understanding of ferroelectric phase behavior in $HfO_2$ but also opens up new avenues for the design and optimization of next-generation semiconductor devices.

Experimental Methods

**Construction of the $HfO_2$ database**

To finetune the MACE model for HfO2, this study employed the on-the-fly learning algorithm[48] integrated with the VASP[49-51] to construct the database. This algorithm autonomously gathers energy, atomic force, and stress data for structures exhibiting significant deviations from those already present in the database during MD simulations. We utilized 2×2×2 supercells 6 $HfO_2$ phases, including $Fm\bar{3}m$, $P2_1/c$, $P4_2nmc$, $Pbca$, $Pca2_1$, and $Pnma$ as the initial structure. All initial structure optimizations were carried out using VASP with the PBEsol exchange-correlation functional. The projector augmented-wave (PAW) method was employed with a plane-wave energy cutoff of 500 eV. The Brillouin zone was sampled using a Γ-centered 4×4×4 k-point mesh for most structures, while a 6×4×3 mesh was used for the Pnma phase. Both atomic positions and lattice parameters were fully relaxed until the maximum force on each atom was below 0.05 eV/Å.

During the on-the-fly learning, the temperature of the system was set from 1 K to 4000 K over 10,000 MD steps under the $NpT$ ensemble, with a time step of 1.5 fs for each phase to cover structures with large variations. Critical structural configurations were subsequently extracted to construct a comprehensive database comprising 4,347 structures. Same as structure optimization, the PBEsol exchange-correlation functional and PAW method was utilized during the on-the-fly learning with a plane-wave cutoff energy of 500 eV, and the same k-point meshes were used.

**Development of the HfO$_2$ MACE model.**

To construct the finetuned MACE model, 80% of the prepared HfO$_2$ dataset was allocated for training, with the remaining 20% used as the test set. Additionally, 5% of the training data was set aside for validation purposes. The model was trained using a batch size of 8 and an initial learning rate of 0.01. Training employed an early stopping criterion with a patience of 5 epochs. During the initial training phase, the exponential moving average (EMA) of the weights was used to stabilize learning. Once the patience threshold was triggered, the optimization strategy shifted to stochastic weight averaging (SWA), which was continued until a total of 100 epochs was reached. The loss function was composed of equally weighted energy and force components, and the AMSGrad variant of the Adam optimizer was adopted to enhance convergence robustness.

**BEC calculation**

The BECs were computed using DFPT[52] as implemented in VASP. This enables the calculation of both the macroscopic dielectric tensor and the BEC tensor from the linear response to finite electric fields. The calculations were performed on fully relaxed structures using the PBEsol exchange-correlation functional with a plane-wave cutoff energy of 500 eV and a Γ-centered 4×4×4 k-point mesh (6×4×3 for the Pnma phase). The ionic contribution to the dielectric response and the BEC tensors were obtained from the linear response to atomic displacements and electric fields.

**NEB calculation**

The minimum energy path of the $Pca2_1$-Pbca and the $P2_1/c$-$Pca2_1$ phase transition was computed using the climbing image nudged elastic band (CI-NEB) method[53] as implemented in VASP. A total of four intermediate images were generated by linear interpolation between the fully relaxed initial and final structures, resulting in five images including the endpoints. A relative spring constant of −5 was used to maintain equal spacing between images. All images were relaxed until the perpendicular forces on the atoms dropped below 0.05 eV/Å. The calculations employed the PBEsol exchange-correlation functional, a plane-wave cutoff energy of 500 eV, and a Γ-centered 4×4×4 k-point mesh. Their total energies without an electric field were computed using the MLP, while the corresponding BECs were evaluated using the BM1 model. The total energy under a specific electric field is then evaluated using the expression proposed by Kutana et al.[21]: $E = E(\varepsilon = 0) - |e| \sum_{\kappa,\alpha\beta} \mathcal{E}_\beta Z^*_{\kappa,\beta\alpha} u_{\kappa,\alpha}$, where $\mathcal{E}_\beta$ is the applied external electric field, $|e|$ is the elementary charge, $Z^*_{\kappa,\beta\alpha}$ is the tensor of BEC, and $u_{\kappa,\alpha}$ is the displacement of ion, which incorporates the coupling between the polarization and the external electric field.

SUPPLEMENTARY MATERIAL

The supplementary material contains the MAE values comparison of energy and force between finetuned MACE and other MLFFs.

AUTHOR INFORMATION

**Corresponding Author**

**Po-Yen Chen** - Department of Materials Engineering, the University of Tokyo, Tokyo, Japan.

**Teruyasu Mizoguchi** - Institute of Industrial Science, the University of Tokyo, Tokyo, Japan.

AUTHOR DECLARATIONS

**Conflict of Interest**

The authors declare no competing financial interest.

**Author Contributions**

**Po-Yen Chen**: Conceptualization; Data Curation; Formal Analysis; Investigation; Methodology; Software; Visualization; Writing/Original Draft Preparation

**Teruyasu Mizoguchi**: Supervision; Funding Acquistion; Writing/Review & Editing

DATA AND CODE AVAILABILITY

The HfO$_2$ structural database and the finetuned HfO$_2$ MACE model have deposited at Zenodo with DOI 10.5281/zenodo.16088425 and are publicly available as of the date of publication. Any additional information required to reanalyze the data reported in this paper is available from the lead contact upon request.

ACKNOWLEDGMENT

This study was supported by the Ministry of Education, Culture, Sports, Science and Technology (MEXT) (Nos. 24H00042), and New Energy and Industrial Technology Development Organization (NEDO). PYC would acknowledge the support of JST SPRING (Grant Number JPMJSP2108).

REFERENCES

1. T. Kang, J. Park, H. Jung, H. Choi, S. Lee, N. Lee, R. Lee, G. Kim, S. Kim, H. Kim, C. Yang, J. Jeon, Y. Kim and S. Lee, High-κ Dielectric (HfO2)/2D Semiconductor (HfSe2) Gate Stack for Low-Power Steep-Switching Computing Devices, *ADVANCED MATERIALS*, 2024, **36**.
2. Y. Yin, C. Huang, P. Chiu, Y. Jiang, J. Hoo and M. Chen, High-Quality HfO2 High-K Gate Dielectrics Deposited on Highly Oriented Pyrolytic Graphite via Enhanced Precursor Atomic Layer Seeding, *ACS APPLIED ELECTRONIC MATERIALS*, 2025, **7**, 1943-1952.
3. D. Lee, Y. Lee, Y. Cho, H. Choi, S. Kim and M. Park, Unveiled Ferroelectricity in Well-Known Non-Ferroelectric Materials and Their Semiconductor Applications, *ADVANCED FUNCTIONAL MATERIALS*, 2023, **33**.
4. L. Baumgarten, T. Szyjka, T. Mittmann, A. Gloskovskii, C. Schlueter, T. Mikolajick, U. Schroeder and M. Müller, Smart Design of Fermi Level Pinning in HfO2-Based Ferroelectric Memories, *ADVANCED FUNCTIONAL MATERIALS*, 2024, **34**.

5.  T. Zhang, Y. Fan, Z. Xue, M. Si, Z. Wang, X. Li and Y. Cao, Coherent epitaxy of HfxZr1-xO2 thin films by high-pressure magnetron sputtering, *MATERIALS TODAY ELECTRONICS*, 2024, **10**.
6.  H. Mulaosmanovic, E. Breyer, S. Dunkel, S. Beyer, T. Mikolajick and S. Slesazeck, Ferroelectric field-effect transistors based on HfO2: a review, *NANOTECHNOLOGY*, 2021, **32**.
7.  P. Fan, Y. Zhang, Q. Yang, J. Jiang, L. Jiang, M. Liao and Y. Zhou, Origin of the intrinsic ferroelectricity of HfO2 from ab initio molecular dynamics, *The Journal of Physical Chemistry C*, 2019, **123**, 21743-21750.
8.  W. Ding, Y. Zhang, L. Tao, Q. Yang and Y. Zhou, The atomic-scale domain wall structure and motion in HfO2-based ferroelectrics: A first-principle study, *ACTA MATERIALIA*, 2020, **196**, 556-564.
9.  J. Wu, Y. Zhang, L. Zhang and S. Liu, Deep learning of accurate force field of ferroelectric HfO2, *PHYSICAL REVIEW B*, 2021, **103**.
10. H. Chen, Y. Zhang, C. Zhou and Y. Zhou, Deep learning potential model of displacement damage in hafnium oxide ferroelectric films, *NPJ COMPUTATIONAL MATERIALS*, 2024, **10**.
11. G. Sivaraman, A. Krishnamoorthy, M. Baur, C. Holm, M. Stan, G. Csanyi, C. Benmore and A. Vazquez-Mayagoitia, Machine-learned interatomic potentials by active learning: amorphous and liquid hafnium dioxide, *NPJ COMPUTATIONAL MATERIALS*, 2020, **6**.
12. P. Friederich, F. Häse, J. Proppe and A. Aspuru-Guzik, Machine-learned potentials for next-generation matter simulations, *NATURE MATERIALS*, 2021, **20**, 750-761.
13. H. Meng, Y. Liu, H. Yu, L. Zhuang and Y. Chu, Machine-learning-potential-driven prediction of high-entropy ceramics with ultra-high melting points, *CELL REPORTS PHYSICAL SCIENCE*, 2025, **6**.
14. J. Zhang, H. Zhang, J. Wu, X. Qian, B. Song, C. Lin, T. Liu and R. Yang, Vacancy-induced phonon localization in boron arsenide using a unified neural network interatomic potential, *CELL REPORTS PHYSICAL SCIENCE*, 2024, **5**.
15. Q. Li, E. Küçükbenli, S. Lam, B. Khaykovich, E. Kaxiras and J. Li, Development of robust neural-network interatomic potential for molten salt, *CELL REPORTS PHYSICAL SCIENCE*, 2021, **2**.
16. C. Lichtensteiger, P. Zubko, M. Stengel, P. Aguado-Puente, J.-M. Triscone, P. Ghosez and J. Junquera, Ferroelectricity in ultrathin film capacitors, *arXiv preprint arXiv:1208.5309*, 2012.
17. P. Ghosez, J. Michenaud and X. Gonze, Dynamical atomic charges:: The case of ABO3 compounds, *PHYSICAL REVIEW B*, 1998, **58**, 6224-6240.
18. K. Shimizu, R. Otsuka, M. Hara, E. Minamitani and S. Watanabe, Prediction of Born effective charges using neural network to study ion migration under electric fields: applications to crystalline and amorphous Li3PO4, *SCIENCE AND TECHNOLOGY OF ADVANCED MATERIALS-METHODS*, 2023, **3**.
19. S. Falletta, A. Cepellotti, A. Johansson, C. Tan, M. Descoteaux, A. Musaelian, C. Owen and B. Kozinsky, Unified differentiable learning of electric response, *NATURE COMMUNICATIONS*, 2025, **16**.
20. L. Ma, J. Wu, T. Zhu, Y. Huang, Q. Lu and S. Liu, Ultrahigh Oxygen Ion Mobility in Ferroelectric Hafnia, *PHYSICAL REVIEW LETTERS*, 2023, **131**.

21. A. Kutana, K. Yoshimochi and R. Asahi, Dielectric tensor of perovskite oxides at finite temperature using equivariant graph neural network potentials, *SCIENCE AND TECHNOLOGY OF ADVANCED MATERIALS-METHODS*, 2025, **5**.
22. A. Kutana, K. Shimizu, S. Watanabe and R. Asahi, Representing Born effective charges with equivariant graph convolutional neural networks, *SCIENTIFIC REPORTS*, 2025, **15**.
23. I. Batatia, P. Benner, Y. Chiang, A. M. Elena, D. P. Kovács, J. Riebesell, X. R. Advincula, M. Asta, M. Avaylon and W. J. Baldwin, A foundation model for atomistic materials chemistry, *arXiv preprint arXiv:2401.00096*, 2023.
24. I. Batatia, D. Kovács, G. Simm, C. Ortner and G. Csányi, 2022.
25. Y. Park, J. Kim, S. Hwang and S. Han, Scalable Parallel Algorithm for Graph Neural Network Interatomic Potentials in Molecular Dynamics Simulations, *JOURNAL OF CHEMICAL THEORY AND COMPUTATION*, 2024, **20**, 4857-4868.
26. H. Yang, C. Hu, Y. Zhou, X. Liu, Y. Shi, J. Li, G. Li, Z. Chen, S. Chen and C. Zeni, Mattersim: A deep learning atomistic model across elements, temperatures and pressures, *arXiv preprint arXiv:2405.04967*, 2024.
27. B. Deng, P. Zhong, K. Jun, J. Riebesell, K. Han, C. Bartel and G. Ceder, CHGNet as a pretrained universal neural network potential for charge-informed atomistic modelling, *NATURE MACHINE INTELLIGENCE*, 2023, **5**, 1031-1041.
28. Y. Yun, P. Buragohain, M. Li, Z. Ahmadi, Y. Zhang, X. Li, H. Wang, J. Li, P. Lu, L. Tao, H. Wang, J. Shield, E. Tsymbal, A. Gruverman and X. Xu, Intrinsic ferroelectricity in Y-doped $HfO_2$ thin films, *NATURE MATERIALS*, 2022, **21**, 903-+.
29. U. Schroeder, T. Mittmann, M. Materano, P. Lomenzo, P. Edgington, Y. Lee, M. Alotaibi, A. West, T. Mikolajick, A. Kersch and J. Jones, Temperature-Dependent Phase Transitions in $Hf_xZr_{1-x}O_2$ Mixed Oxides: Indications of a Proper Ferroelectric Material, *ADVANCED ELECTRONIC MATERIALS*, 2022, **8**.
30. Y. Tashiro, T. Shimizu, T. Mimura and H. Funakubo, Comprehensive Study on the Kinetic Formation of the Orthorhombic Ferroelectric Phase in Epitaxial Y-Doped Ferroelectric $HfO_2$ Thin Films, *ACS APPLIED ELECTRONIC MATERIALS*, 2021, **3**, 3123-3130.
31. H. Jónsson, G. Mills and K. W. Jacobsen, in *Classical and quantum dynamics in condensed phase simulations*, World Scientific, 1998, pp. 385-404.
32. T. Zhu, L. Ma, S. Deng and S. Liu, Progress in computational understanding of ferroelectric mechanisms in $HfO_2$, *NPJ COMPUTATIONAL MATERIALS*, 2024, **10**.
33. D. Choe, S. Kim, T. Moon, S. Jo, H. Bae, S. Nam, Y. Lee and J. Heo, Unexpectedly low barrier of ferroelectric switching in $HfO_2$ via topological domain walls, *MATERIALS TODAY*, 2021, **50**, 8-15.
34. Y. Wu, Y. Zhang, J. Jiang, L. Jiang, M. Tang, Y. Zhou, M. Liao, Q. Yang and E. Tsymbal, Unconventional Polarization-Switching Mechanism in $(Hf, Zr)O_2$ Ferroelectrics and Its Implications, *PHYSICAL REVIEW LETTERS*, 2023, **131**.
35. R. Batra, T. Huan, J. Jones, G. Rossetti and R. Ramprasad, Factors Favoring Ferroelectricity in Hafnia: A First-Principles Computational Study, *JOURNAL OF PHYSICAL CHEMISTRY C*, 2017, **121**, 4139-4145.
36. M. Kingsland, S. Lisenkov, S. Najmaei and I. Ponomareva, Phase transitions in $HfO_2$ probed by first-principles computations, *JOURNAL OF APPLIED PHYSICS*, 2024, **135**.
37. K. Hisama, G. Huerta and M. Koyama, Molecular dynamics of electric-field driven ionic systems using a universal neural-network potential, *COMPUTATIONAL MATERIALS SCIENCE*, 2023, **218**.

38. R. Han, P. Hong, S. Ning, Q. Xu, M. Bai, J. Zhou, K. Li, F. Liu, F. Shi, F. Luo and Z. Huo, The effect of stress on HfO2-based ferroelectric thin films: A review of recent advances, *JOURNAL OF APPLIED PHYSICS*, 2023, **133**.
39. K. Ye, T. Jeong, S. Yoon, D. Kim, Y. Kim, C. Hwang and J. Choi, Ab Initio Study on 3D Anisotropic Ferroelectric Switching Mechanism and Coercive Field in HfO2 and ZrO2, *ADVANCED FUNCTIONAL MATERIALS*, 2025.
40. H. Cheng, P. Jiao, J. Wang, M. Qing, Y. Deng, J. Liu, L. Bellaiche, D. Wu and Y. Yang, Tunable and parabolic piezoelectricity in hafnia under epitaxial strain, *NATURE COMMUNICATIONS*, 2024, **15**.
41. S. Liu and B. Hanrahan, Effects of growth orientations and epitaxial strains on phase stability of HfO2 thin films, *PHYSICAL REVIEW MATERIALS*, 2019, **3**.
42. S. Lee, H. Jeong, J. Park, D. Lee, Y. Jo, J. Yang, M. Park and S. Lee, Enhanced ferroelectric performance in Hf0.5Zr0.5O2 capacitors using ultra-thin MoS2 layer for clamping effect and oxygen vacancy suppression, *MATERIALS & DESIGN*, 2025, **254**.
43. S. Kim, D. Narayan, J. Lee, J. Mohan, J. Lee, J. Lee, H. Kim, Y. Byun, A. Lucero, C. Young, S. Summerfelt, T. San, L. Colombo and J. Kim, Large ferroelectric polarization of TiN/Hf0.5Zr0.5O2 capacitors due to stress-induced crystallization at low budget, *APPLIED PHYSICS LETTERS*, 2017, **111**.
44. G.-W. Liu, W. Zaheer, L. Carrillo and S. Banerjee, Metastable polar orthorhombic local structure of hydrothermally grown HfO2 nanocrystals, *Cell Reports Physical Science*, 2024, **5**.
45. S. Estandía, N. Dix, J. Gazquez, I. Fina, J. Lyu, M. Chisholm, J. Fontcuberta and F. Sánchez, Engineering Ferroelectric Hf0.5Zr0.5O2 Thin Films by Epitaxial Stress, *ACS APPLIED ELECTRONIC MATERIALS*, 2019, **1**, 1449-1457.
46. Z. Gao, S. Lyu and H. Lyu, Frequency dependence on polarization switching measurement in ferroelectric capacitors, *JOURNAL OF SEMICONDUCTORS*, 2022, **43**.
47. Y. Li, J. Li, R. Liang, R. Zhao, B. Xiong, H. Liu, H. Tian, Y. Yang and T. Ren, Switching dynamics of ferroelectric HfO2-ZrO2 with various ZrO2 contents, *APPLIED PHYSICS LETTERS*, 2019, **114**.
48. R. Jinnouchi, F. Karsai and G. Kresse, On-the-fly machine learning force field generation: Application to melting points, *PHYSICAL REVIEW B*, 2019, **100**, 014105.
49. G. Kresse and J. Furthmuller, Efficient iterative schemes for ab initio total-energy calculations using a plane-wave basis set, *PHYSICAL REVIEW B*, 1996, **54**, 11169-11186.
50. G. KRESSE and J. HAFNER, AB-INITIO MOLECULAR-DYNAMICS SIMULATION OF THE LIQUID-METAL AMORPHOUS-SEMICONDUCTOR TRANSITION IN GERMANIUM, *PHYSICAL REVIEW B*, 1994, **49**, 14251-14269.
51. G. Kresse and J. Furthmuller, Efficiency of ab-initio total energy calculations for metals and semiconductors using a plane-wave basis set, *COMPUTATIONAL MATERIALS SCIENCE*, 1996, **6**, 15-50.
52. X. Gonze and C. Lee, Dynamical matrices, born effective charges, dielectric permittivity tensors, and interatomic force constants from density-functional perturbation theory, *PHYSICAL REVIEW B*, 1997, **55**, 10355-10368.
53. G. Henkelman, B. Uberuaga and H. Jónsson, A climbing image nudged elastic band method for finding saddle points and minimum energy paths, *JOURNAL OF CHEMICAL PHYSICS*, 2000, **113**, 9901-9904.

Tables

| MAE | Finetuned MACE | MACE-MP-0 | SevenNet-0 | MatterSim | CHGNET |
|---|---|---|---|---|---|
| Energy (meV/atom) | **5.01** | 35.68 | 26.84 | 23.06 | 41.60 |
| Force (meV/Å) | **34.09** | 172.80 | 135.05 | 75.67 | 156.73 |

**Table 1.** MAE values comparison of energy and atomic force between the finetuned MACE and other MLFFs[23, 25-27].

Figures

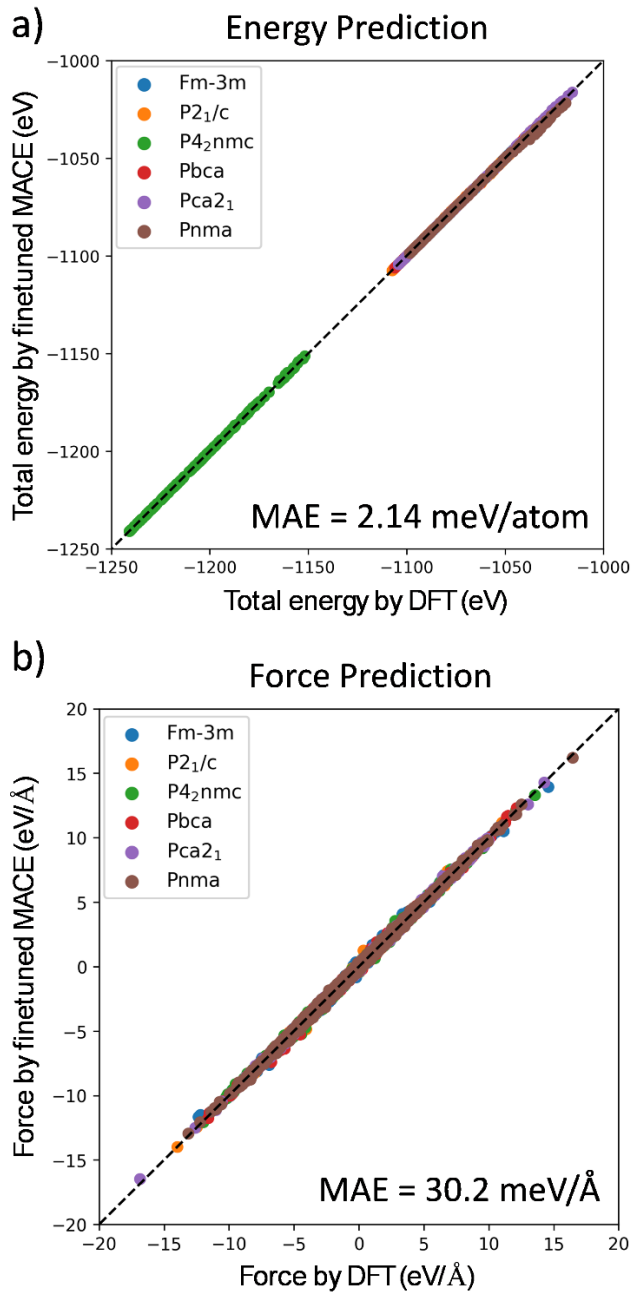

**Figure 1**. The a) total energy and b) force comparison between finetuned MACE and DFT calculation for 6 $HfO_2$ phases.

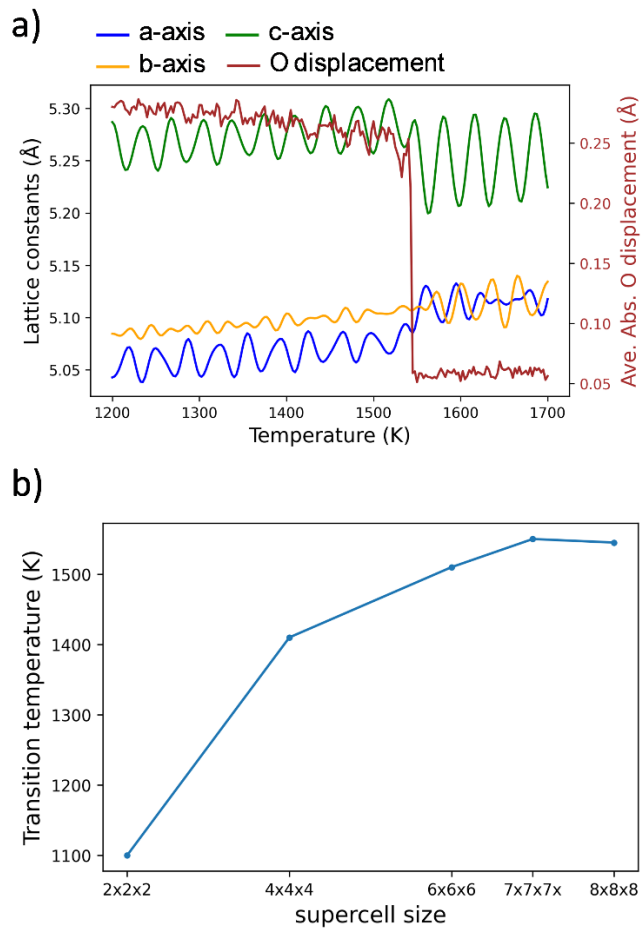

**Figure 2**. a) The temperature-dependent lattice constant (blue: a axis, orange: b axis, green: c axis) and averaged absolute Pca2$_1$ displacement along [010] direction for the 8×8×8 supercell. b) The supercell-size dependent transition temperature of the Pca2$_1$ to P4$_2$nmc phase transition.

**Figure 3**. The prediction comparison of BECs between the pretrained BM1 model and DFPT calculations for 6 $HfO_2$ phases.

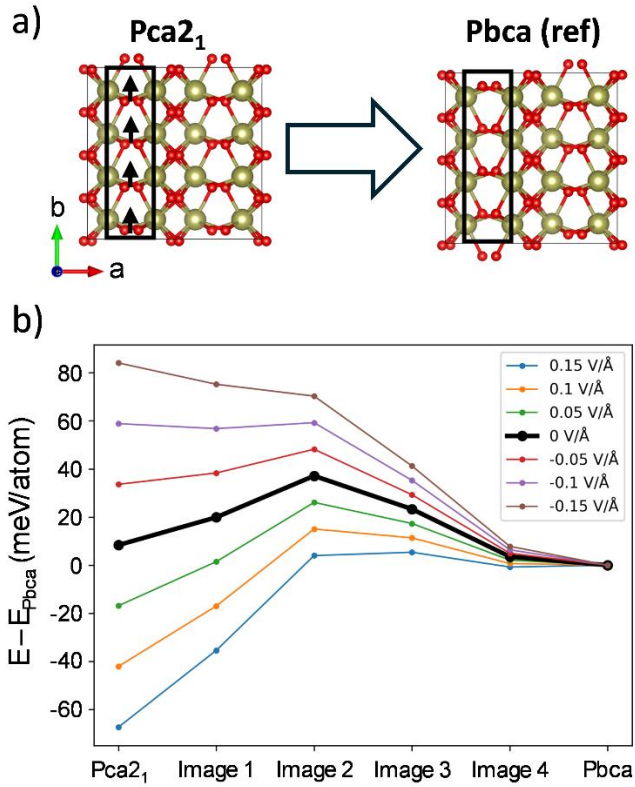

**Figure 4**. a) The scheme of the Pca2$_1$→ Pbca phase transition, where the arrow shows the oxygen movement, and the b) its NEB diagram under various electric fields along the [010] direction.

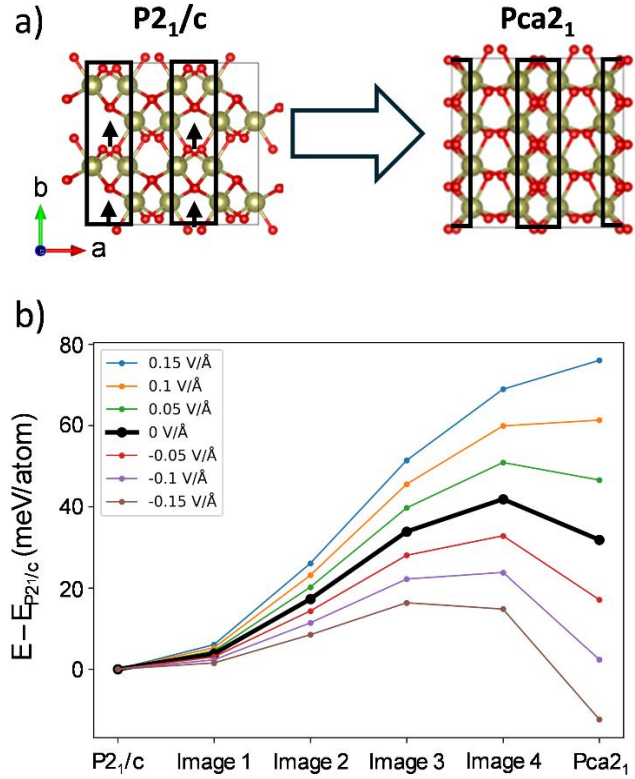

**Figure 5**. The a) scheme of the P2$_1$/c → Pca2$_1$ phase transition, where the arrow shows the oxygen movement, and the b) its NEB diagram under various electric fields along the [010] direction.

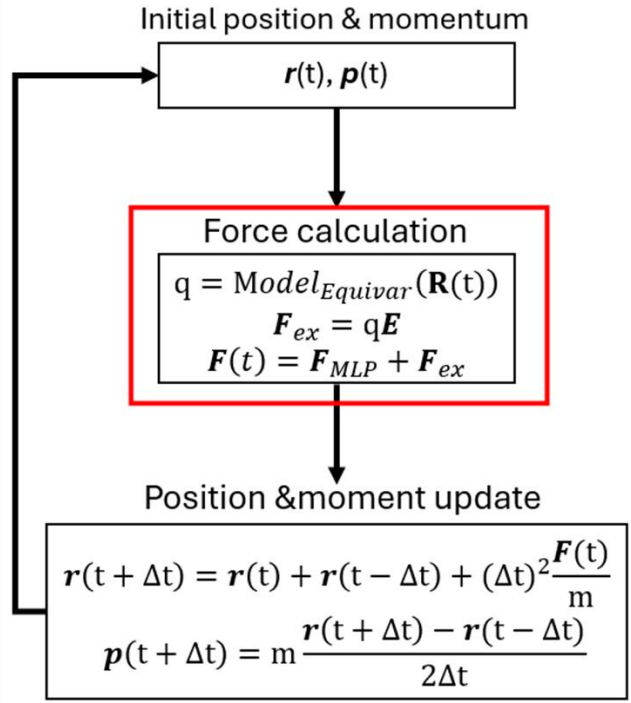

**Figure 6**. The flow chart for iteration during MD simulation with BECs under an external electric field.

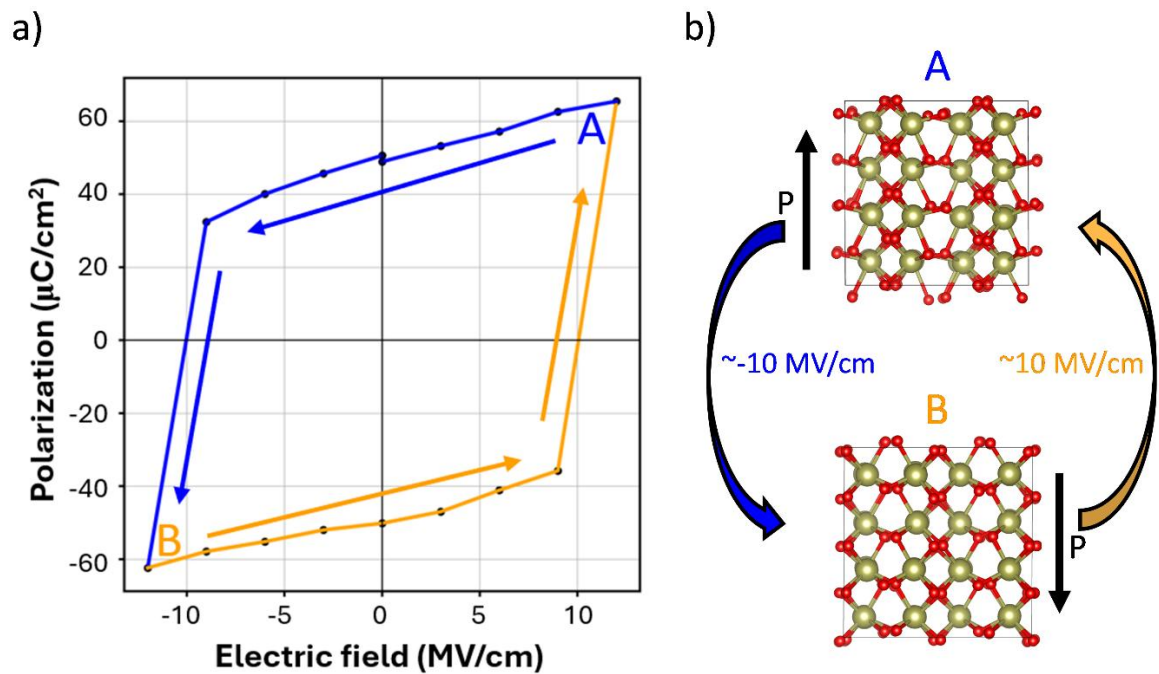

**Figure 7**. a) The hysteresis loop of Pca2$_1$ HfO$_2$ obtained by the MLP-based MD simulation under electric fields and b) the scheme of polarization switching in the Pca2$_1$ phase.

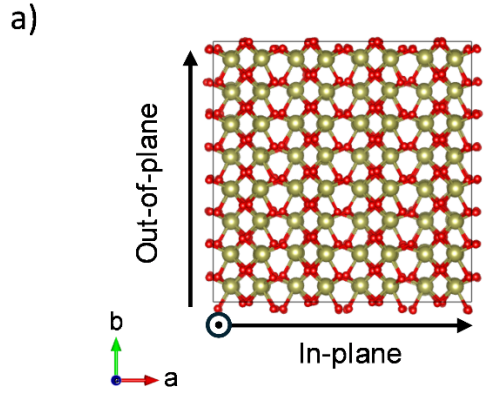

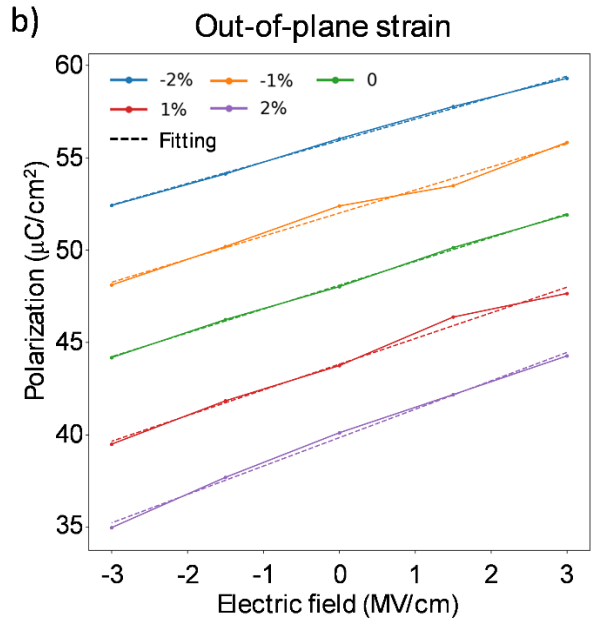

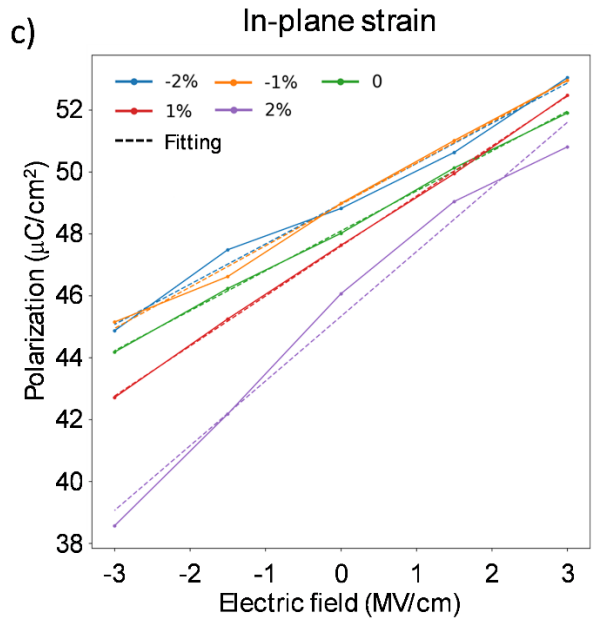

**Figure 8**. a) The scheme for the definition of out-of-plane and in-plane strain, the P-E curve of HfO$_2$ under various b) out-of-plane strains and c) in-plane strains. The point line represents the simulated data, and the dotted line indicates the linear regression fit.

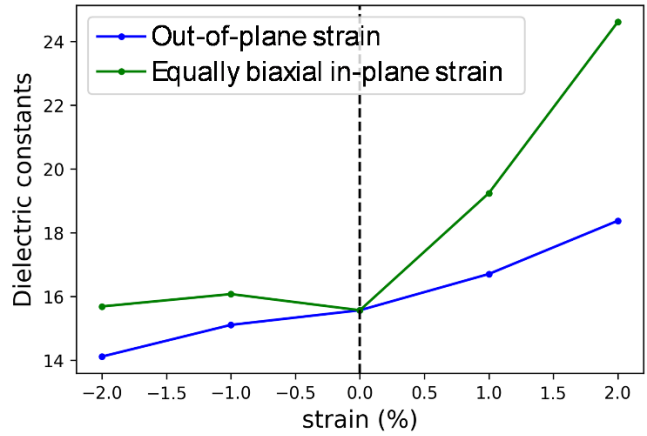

**Figure 9**. The strain-dependent dielectric constant for (blue) out-of-plane strain and (green) in-plane strain.

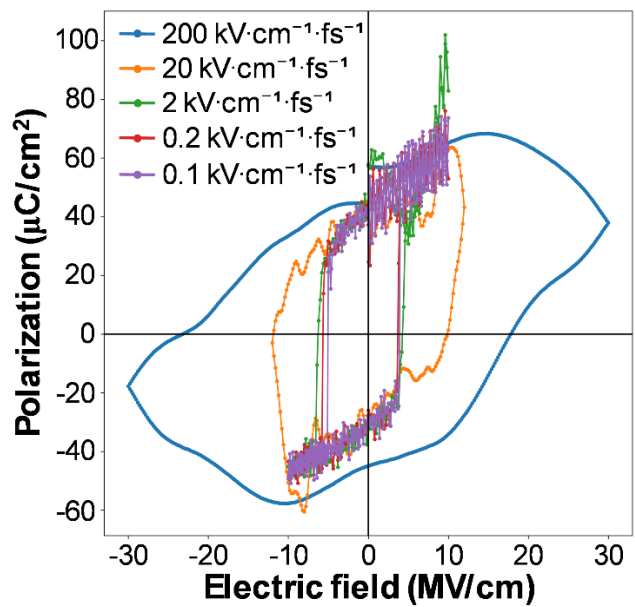

**Figure 10**. The P-E hysteresis loop with the triangular-wave electric field of the changing rate of 200 (blue), 20 (orange), 2 (green), 0.2 (red), and 0.1 (purple) kV·cm$^{-1}$·fs$^{-1}$.